# Nonlinear behavior of vibrating molecules on suspended graphene waveguides

Amrita Banerjee and Haim Grebel(*)

*Electronic Imaging Center and Electrical Engineering Department, New Jersey Institute of Technology, Newark, NJ 07102. USA*

**Abstract:**

Suspended graphene waveguides were deposited on micron-scale periodic metal (plasmonic) structures. Raman scattering of test molecules (*B. Megaterium*), deposited on the waveguides' surface, exhibited azimuthal cycles upon rotation: at these micron scales, spontaneous Raman ought to be independent of phase matching conditions. In addition, we observed angular-selective quadratic intensity dependence contrary to the typical linear behavior of spontaneous Raman. The effects were observed at very modest pump laser intensities (<10 MW/cm$^2$ at the sample surface, oftenly used in Raman experiments). We attributed these observations to nonlinear coupling between the vibrating molecules and surface plasmon polariton (SPP) modes at the molecular vibration frequency. It was assessed that the polariton mode propagates through fairly long distances (over 100 microns).

(*) Corresponding author:grebel@njit.edu

Graphene [1] has attracted much attention lately for the least of exhibiting low loss in the THz/infrared range [2]. Applications at sub THz/infrared frequency range [3-4], and the combination of graphene with resonating meta-structures at near IR wavelengths ($\lambda \sim 1600$ nm) [5] have been introduced, as well. It has also been known that when interfaced with resonating periodic metallic structures, graphene enables better linear and nonlinear interaction with molecules at its surface [6]. Here we address the nonlinear interaction between surface plasmon polariton (SPP) waves in mid-IR ($\lambda$ in the range of 6 to 10 μm), that are propagating in suspended graphene waveguides, and local vibrating molecules, situated at the waveguide surface (Fig. 1).

Plasmonic structures (metal mesh screens) have been studied in the short and long wavelength regions for astronomy, chemistry and biology purposes [7]. These screens enable the separation of desired IR signals from more energetic short wavelength radiation, allow for color temperature measurements, provide order sorting for grating spectrometers and improve the signal-to-noise ratio of Fourier transform infrared spectrometers (FTIR) [8-11]. Periodic metal/dielectric structures exhibit resonances when the propagating surface wavelength is on the order of the characteristic screen's pitch. Additional effects may be observed due to resonating characteristics (if present) of the screen's individual features, similarly to photonic crystals. In the far-field, the screens transmit or reflect particular frequency bands, which qualify them as optical filters. Here, we are interested in near-field nonlinear interactions between molecules and SPP waves in the mid-IR range.

Nonlinear interaction between molecular vibrations and SPP modes is the subject of surface enhanced Raman scatterings (SERS) [12 and references therein]. The near-field aspects of the pump and scattered beams dictate usage of sub-micron metal structures (colloids, rod shaped optical-antennas, etc.,); in contrast, SPP modes propagating at phonon frequencies require metal structures on the scale of tens of micrometers. Study of SPP modes at phonon frequencies (mid-IR and THz frequencies) has been less prevalent perhaps for lack of appropriate waveguides.

**Experiment and Results:** Freestanding, square shaped, electroformed copper and nickel screens were purchased from Precision Eforming. The 1x1 cm$^2$ screens were 4 microns thick. Their periodicities were 12.6 μm and 12 μm for copper and nickel, respectively. The screen's opening was 7.6x7.6 μm$^2$. These screens are considered thick, implying higher order of diffraction and the formation of waveguide modes within the structure opening. Highly ordered pyrolithic graphite (HOPG) was purchased from SPI Supplies and deposited on the screen's surface as per our previous recipe [13]. Such procedure results in a few layered graphene, typically 3-4 layers as assessed by Raman spectroscopy. In addition, the 2700 cm$^{-1}$ line of graphene has substantially narrowed when placed on these periodic screen structures, alluding to the generation of phonon/polariton surface waves in graphene-coated metal-mesh substrates. Spores of *B. Megaterium* were suspended in ethanol and a few drops were used directly on the substrate. Raman spectra were taken with a 514.5 nm line of Ar ion laser at a typical 10 mW/μm$^2$ at the sample when focused by x50 extended focal length objective (N.A=0.55) or x40 objective (N.A=0.65), a 75 cm spectrometer equipped with a 1200 g/mm grating (resolution of 1 cm$^{-1}$) and a cooled CCD detector array. The laser was focused at the center of the screen's opening, far from any scattering edge. The sample holder had a large opening through which the shadow of the screen was visible when the laser was focused to a point above the transparent graphene. The laser spot was much smaller than either the screen's pitch or its opening. Ten successive measurements of 10 seconds each were taken; dark current background was subtracted from the data, yet, the overall background signal was left without modification.

Scanning electron microscope picture of graphene deposited on top of the metal screen is shown in Fig. 1a and the experimental configuration is shown in Fig. 1b. The pump laser was linearly polarized within the plane of incidence (p-polarization). The Raman spectrum of *B. Megaterium* is shown in Figure 1c. The spectrum is similar to what has been reported in the past [14].

Peak intensity at 900 cm$^{-1}$ as a function of in-plane rotation angle is shown in Fig. 2a. The peak intensity exhibited the four-fold symmetry of the screen. A control experiment on screens, coated with just the graphene layer, did not exhibit such effect at 900 cm$^{-1}$. Another control experiment with the *B. Megaterium* on a screen, without the graphene layer, did not exhibit such symmetry either. It was easy to verify that the interrogating spot did not wobble upon screen's rotations; the graphene is transparent and the screen's shadow was visible below the screen's holder. The shadow of the screen helped us establishing a reference (initial) point for the in-plane rotation angle data.

Raman signal as a function of tilt angle is shown in Fig. 2b. If we to consider the spread in laser spot size as a function of tilt angle (and the related reduction in the Raman signal as a result of lower laser intensity), this would behave as $cos\theta$: the Raman signal at $\theta=20°$ would be 94% of the peak power.

In general, at small pump intensities, the Raman signal depends on the laser pump intensity $I_L$ linearly. Yet, intensity measurements (Figure 2c,d) suggested an additional $I_L^2$ term at specific tilt angles. In Figure 2c,d we added the Raman line for the graphene waveguides as a reference. The graphene line was measured with the supported *B. Megaterium* at its surface.

Yet, when the platforms were tilted to $\theta=24°$, the graphene exhibited the quadratic intensity term while the *B. Meg.* exhibited only a linear behavior.

**Discussion and Analysis:** Raman data exhibited in-plane rotation symmetry seen only for periodic structures at pump wavelength scales (sub-micron scales) [15]. This is particularly puzzling because the laser spot is small and is focused at the center of the screen's opening, far from any scattering edge. At micron scales there should not be coupling between the laser frequency and the structure. So far, angular dependence at micron scale was exhibited by (the linear) infrared spectroscopy [11] as a result of resonance coupling between the infrared radiation and the periodic screen. We also observed angular selective quadratic intensity dependence for two difference molecules situated on the same substrate whereas typically, spontaneous Raman depends on the pump intensity linearly. These, seemingly unrelated observations may be attributed to the formation of SPP modes at the vibrating frequencies. Such surface modes are polarized perpendicularly to the surface, are sensitive to the symmetry of plasmonic structures at the micron scale and when coupled back to the vibrating molecules contribute to additional nonlinear Raman signal.

Thermal effects and line selective intensity dependence could be ruled out in our case: the quadratic effect was seen only upon tilting the samples to specific angles, which as we shall see later, may be exactly calculated. The graphene membrane seems essential; without it we found no azimuthal cycling symmetry. The type of metal (copper, nickel) did not play any role in our measurements, either. Thinner films produced better results; dried-out samples, flaking off the substrates, were not appropriate for the measurements.

Infrared p-polarized radiation is best coupled to the polariton surface mode propagating in the graphene waveguide through momentum conservation (similarly to [16]):

$$sin(\theta) = |G'|cos(\phi) \pm \sqrt{(n_{eff})^2 - |G'|^2(1-cos^2(\phi))}. \qquad (1)$$

Here: $\phi$ is the in-plane rotation angle between the projection of the incident wavevector $\boldsymbol{k}_0$ on the screen and $\boldsymbol{G}$; $\boldsymbol{G'}=\boldsymbol{G}/k_0$ with $\boldsymbol{G} = \hat{x}q_1\frac{2\pi}{a} + \hat{y}q_2\frac{2\pi}{a}$, the reciprocal wavevector of the lattice of screen's openings with a pitch $a$. The propagating SPP wavelength is calculated as, $\lambda = 1/\bar{k}_{phonon}$. Here, $\bar{k}_{phonon}$ is the Raman shift frequency. For example, a 900 cm$^{-1}$ Raman shift peak of *B. Megaterium* is equivalent to $\lambda$=11.1 µm. If propagation is made along the x-direction of the screen, then, $q_1$=1 and $q_2$=0. For a waveguide loaded with air, $n_{eff}\cong$1 [16]. In our case $a$=12 µm; for $\phi\sim\pi/2$, the optimal tilt angle of the screen with respect to the incident infrared beam would be $\theta$=4.3°. For graphene, $\lambda\sim$1/(1600 cm$^{-1}$)=6.25 µm and $\theta$=28.6°. Standing surface waves are invoked by the same periodic structure, yet not necessarily along the same direction, thus forming resonance conditions [6].

The symmetry of the metal mesh screen dictates the efficiency of the coupling process, as well. Thick screens with square lattice of opening exhibit four-fold symmetry and the formation of symmetric and anti-symmetric modes [17]. Thus, as we rotate the samples in-plane, the electric field mode distribution should behave as $cos(4\phi)$ and its intensity profile as $cos^2(4\phi)$. Namely, we will observe 45° azimuthal cycles. This is quite distinct from the Raman signal of say, aligned carbon nanotubes, which exhibited 180° azimuthal cycles [18].

Once excited, these long-wavelength SPP modes may form standing waves within the x-y plane through Bragg scatterings, $2\beta_{spp}cos(\phi')\sim mG$ where $\phi'$ is the angle between $G$ and $\beta_{spp}$ with $m$ integer. In our case, $\beta_{spp}\sim k_{\Omega}(=2\pi/11.1)n_{eff}(\approx 1)>G(=2\pi/12)$ µm$^{-1}$ and thus $\phi'=22°$ for $m=2$. A plot of $\theta$ as a function of incident $\phi$ for a screen with pitch $a=12$ µm and incident wavelength of 11.1 µm is given in Fig. 3. The tilt angle is fairly constant for a wide range of in-plane rotation angles and one can satisfy both the Bragg condition for a standing wave formation and the condition for optimal coupling to an SPP mode. Based on the fit of Fig. 2a we find that $\phi_o=12°$ was satisfying both requirements.

Raman interaction involves three frequencies: the pump at $\omega_L$, the scattered frequency at $\omega_S$ and the vibrating molecule's frequency at $\Omega$; typically, $\Omega\sim\omega_L-\omega_S$. With $X$ the molecular displacement in normal vibration coordinates, $E$ the local electric field, $\alpha$ the polarizability tensor and $N$ is the number of molecules involved,. the nonlinear polarization of a scattered wave is written as, $P_{NL}(\omega_S)=\varepsilon_0 N(\partial\alpha/\partial X)_0:XE=[\delta\varepsilon_0]exp(i\Omega t)E(\omega_L)exp(i\omega_L t)+cc$, with $[\delta\varepsilon_0]$, the dielectric tensor [19]. Yet, scattering into a propagating polariton mode at frequency $\Omega$ requires a high-order term: $P_{NL}(\Omega)\sim[\delta\varepsilon_1]exp(i\Omega t)E(\omega_L)\cdot E^*(\omega_L)+cc$. This would constitute a phonon/polariton mode. Scattering into a propagating SPP mode may require non-inversion symmetry, which could be provided by the thin graphene layer.

We use the new nonlinear polarization term as source to the wave equation for the propagating SPP mode, $\delta(x,y)\Omega^2\mu_0 P_{NL}(\Omega)$. This would be a point source because the laser spot $d$ is much smaller than the propagating SPP wavelength, the screen's pitch or its opening. Its polarization would depend on the polarization of the incident laser beam. The solution everywhere (except for the origin) is simply a combination of plane waves, $E(\Omega,z)=E_0(\omega_L,\omega_S,\Omega)exp[-i(\Omega t-ik_{\Omega}x)]+c.c$ with a mid-IR wavelength $\lambda=2\pi/k_{\Omega}$. The propagation direction will be dictated by optimal coupling to a surface mode as argued earlier, say along the x-direction.

The analysis of standing surface wave formation may use coupled-mode theory for the generated forward and backward SPP waves [20]. The intensity reflection coefficient is written as, $R = |E_\Omega^+ / E_\Omega^-|^2 = \frac{\kappa^2 sh^2(s \cdot L)}{\kappa^2 ch^2(s \cdot L) + (\frac{\Delta\beta}{2})^2 sh^2(s \cdot L)}$, where $L$ is the interaction length along the direction of propagation of the SPP mode, $s^2 = \kappa^2 - (\frac{\Delta\beta}{2})^2$, $\kappa$ is the coupling constant between the forward $E_\Omega^+$ and the backward $E_\Omega^-$ propagating SPP modes, and $\Delta\beta = k_\Omega^+ - k_\Omega^- - qG = 2[k_\Omega(sin(\theta)) - k_\Omega(sin(\theta_0))]$ is the phase mismatch between the two. The red fit curve for $R$ as a function of $\theta$ in Fig. 2b is based on the assumption that the Raman signal is sensitive to such counter propagating and reflection process.

The reflected SPP mode is coupled back to the molecules and contributes to the scattered wave at $\omega_S$. The reflection coefficient $R$ may assess the formation of standing surface modes. We assume for simplicity that the SPP propagates along the x-direction. Using slowly varying envelope approximation, the Raman signal developed within a spot of diameter $d<<L$ is,

$$\frac{\partial}{\partial x} E_S = i\frac{\omega_S^2 \mu_0}{2k_S}[\boldsymbol{P}_{NL}^{(3)}(\omega_S) + \boldsymbol{P}_{NL}^{(5)}(\omega_S)]exp(-ik_S x) =$$
$$= i\frac{3\omega_S^2}{4k_S c^2}[\chi^{(3)} : |\boldsymbol{E}_L|^2 \boldsymbol{E}_S + \frac{5}{4}\chi^{(5)} : r E_\Omega^+ \boldsymbol{E}_L exp(i\Delta k x)]$$

(2)

Here, $r = E_\Omega^- / E_\Omega^+$, the reflection coefficient induced on the SPP mode by the periodic structure. The phase mismatch in the exponent is $\Delta k = k_\Omega - G + k_L - k_S$. The wavevectors $k_L$, $k_S$ have small projection on the x-y plane. Phase matching is introduced through $k_\Omega \sim G$, otherwise, the efficiency of that term is substantially reduced. The last term in Eq. 2 also suggests that the scattered frequency $\omega_S$ is the results of two-wave difference mixing between the laser beam and the SPP mode. Since, $E_\Omega \sim E_S |E_L|^2 |E_L|^2$, we essentially deal with a stimulated Raman process, or fifth-order nonlinearity. The intensity of the scattered signal at $\omega_S$ is,

$$I_S \sim I_S(0) exp(\frac{i3\omega_s \chi^{(3)} I_L d}{2\varepsilon_0 c^2 n_s n_L} + \frac{i15\omega_s \chi^{(5)} R I_L^2 d}{8\varepsilon_0 c^3 n_s n_L^2}). \qquad (3)$$

Note that the polarization effect, and hence the sensitivity to azimuthal rotations, originates from the last tern in (2) when $k_\Omega \sim G$. For relatively small pump intensities, the exponent may be expanded to first-order of approximation and we find that, at resonance, the Raman signal exhibits an added nonlinear term behaving as $RI_L^2$ (Fig 2). Thus, the intensity dependence is attributed to the multipath of a resonating SPP mode along the graphene waveguide.

The optimal tilt angle for the graphene line at 1600 cm$^{-1}$ ($\lambda = 1/\bar{k}_{phonon} = 6.25 \mu m$) would be $\theta=28.6°$ (Eq. 1), far from $\theta \sim 4°$ (the resonance angle for the 900 cm$^{-1}$ line). The fit in Fig. 2b was based on the assumption that the phase mismatch is $\Delta\beta=k_\Omega(\sin\theta-\sin\theta_0)$ and with a constant reflectivity background of $R_0=0.4$. The interaction length was taken as, $L\sim 50$ μ and $\kappa\sim 0.025$ μm$^{-1}$. The results allude to the long range of phonon/polariton propagation constituting more than 4 lattice constants (or 8 lattice constants back and forth).

**Conclusions:** Suspended graphene waveguides over micron-scale metal-mesh screens were used in a study of Raman scattering. Raman signals of a test molecule *B. Megaterium* were found sensitive to in-plane rotations and tilt of the waveguides with respect to the incident, linearly polarized beam. When at plasmonic resonance, the Raman signal exhibited an additional quadratic effect. Overall, graphene is a good waveguide for mid-infrared applications.

**Acknowledgement:** The research was partially funded by NSF IIS-0514361. Support by the National MASINT Management Office (NMMO) is greatly appreciated. We also thank D. Hahn of TX State University for providing us with the *B. Megaterium*.

**Figure Captions**

Fig. 1: (a) Scanning electron microscope picture of graphene-coated metal screen. The laser beam was focused at the opening center, and the substrate was tilted and rotated as necessary. (b) Configuration of the experiment. (c) Raman spectra of *B. Megaterium*.

Fig. 2: (a) Raman signal of *B. Megaterium* at 900 cm$^{-1}$ as a function of in-plane rotation angle for $\theta=4°$ (resonance condition for the 900 cm$^{-1}$ line). The graphene coated a nickel substrate. The best fit curve (light blue curve) was made with $y_0+A\sin^2(2\pi\cdot\phi/B+\phi_0)$; $B=94.7°$ (a cycle of ~4) and $\phi_0=1.8$ rad. Control experiments for graphene-only coatings (G-IR) and *B. Meg* on screens without graphene did not exhibit such symmetry. (b) Blue dots: experimental data points for Raman signal as a function of tilt angle at $\phi\sim180°$. The values were normalized by highest value. Solid red curve: in-plane reflection coefficient as a function of tilt angle (in degrees). Here, $\kappa=0.025$ μm$^{-1}$; the minimum interaction length was taken as $L\sim50$ μm; resonance tilt angle was $\theta_0=4°$ and the background reflection was taken as $R_0=0.4$. (c) Raman signal as a function of pump intensity for *B. Meg.* at 900 cm$^{-1}$ and for graphene at 1600 cm$^{-1}$ (same sample). The sample orientation was $\theta_0=4°$ and $\phi=180°$. The nonlinear fit as a function of laser intensity for *B. Meg* was based on $a+bI_L+cI_L^2$ while that for graphene was fitted with a linear curve. The actual pump intensity values at the sample' surface were 30% of the values given for the axis due to mirrors and beam splitter losses. (d) Roles reversed at $\theta=24°$ and $\phi=180°$. We were limited by the largest tilt angle available with the x40 objective.

Fig. 3.  Simulations of Eq. 1: tilt angle as a function of in-plane rotation angle (in degrees) for $\lambda=11.1$ μm and pitch of 12 μm, respectively.

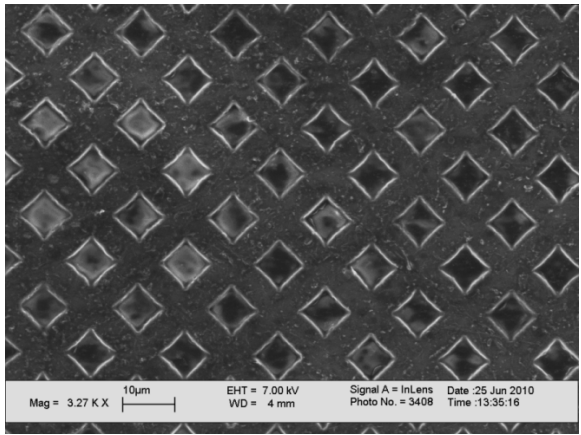

(a)

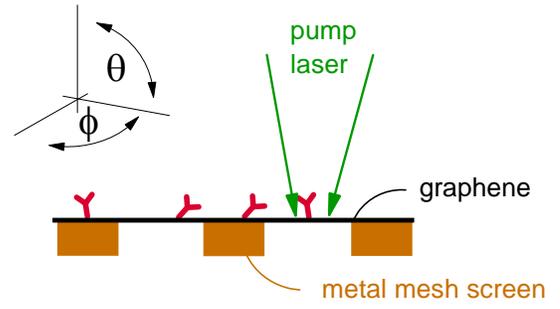

(b)

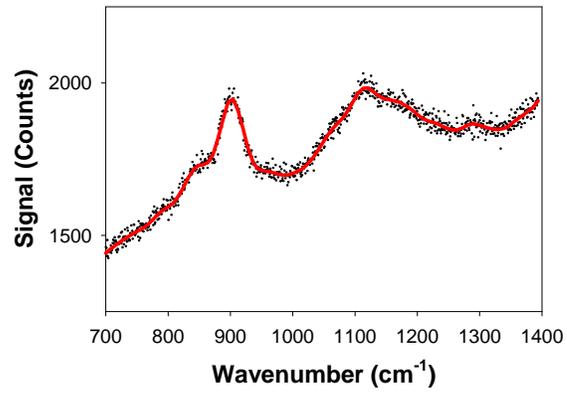

(c)

Fig. 1.

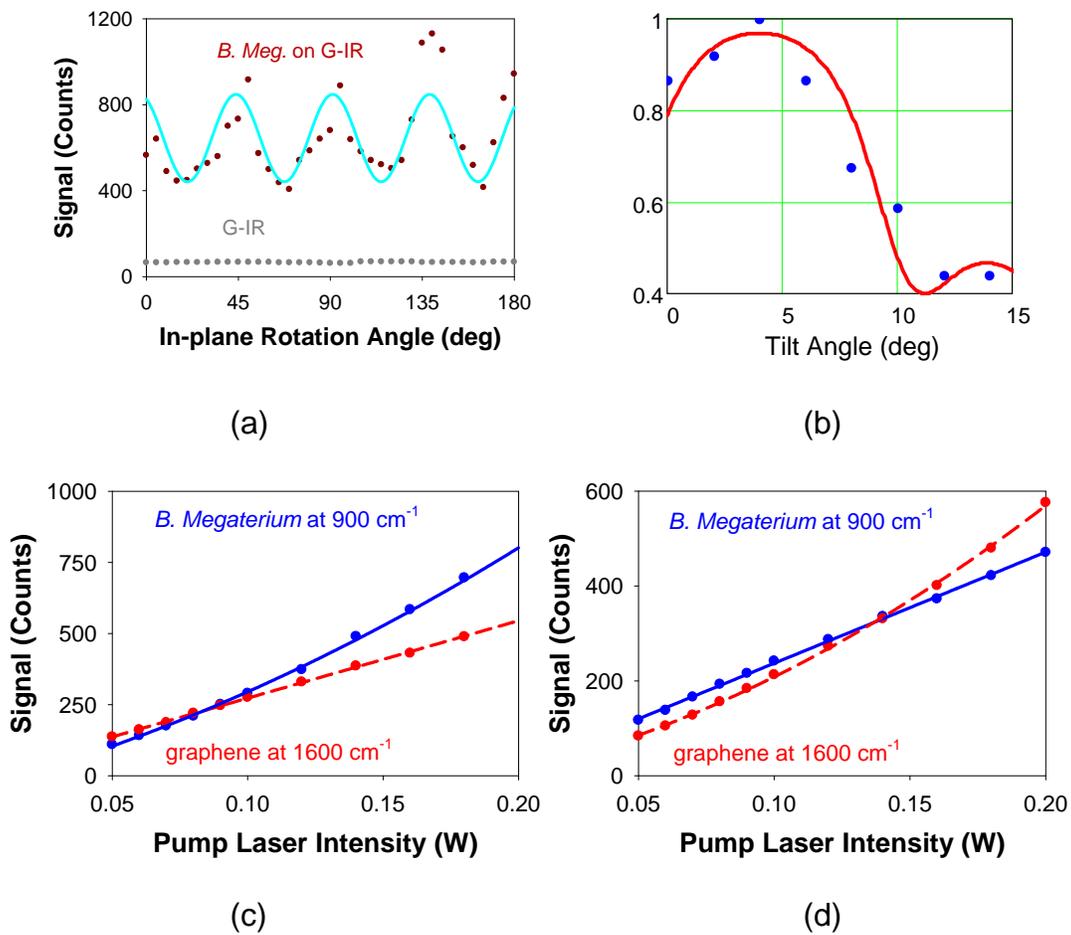

Fig. 2.

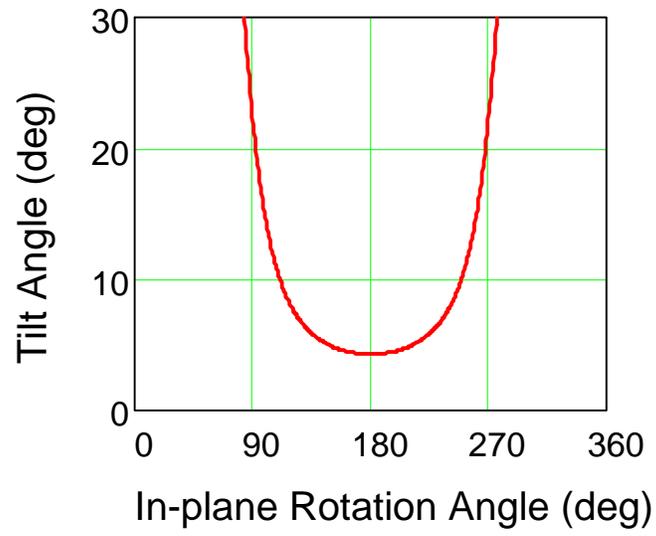

Fig. 3.